\documentclass[10pt,letterpaper,twocolumn]{article}
\usepackage[draft]{hyperref}

\usepackage{ol2}
\usepackage{amsmath}
\usepackage{amssymb}
\usepackage{color}

\begin{document}

\twocolumn[

\title{Generation of indistinguishable and pure heralded single photons with tunable bandwidth}

\author{Xiaojuan Shi, Alejandra Valencia, Martin Hendrych, and Juan P. Torres}

\address{ICFO-Institut de Ciencies Fotoniques, and
Department of Signal Theory and Communications, Universitat
Politecnica de Catalunya, Castelldefels, 08860 Barcelona, Spain}
\email{juan.perez@icfo.es}

\begin{abstract}
We describe a new scheme to {\em fully} control the joint spectrum
of paired photons generated in spontaneous parametric
down-conversion. We show the capability of this method to generate
frequency-uncorrelated photon pairs that are pure and indistinguishable,
and whose bandwidth can be readily tuned. Importantly, the scheme
we propose here can be implemented in any nonlinear crystal and
frequency band of interest.
\end{abstract}

\ocis{270.0270, 270.5585, 270.5565, 190.4410}] \maketitle

The generation of pure and indistinguishable single photons with a
well defined spatio-temporal mode is a fundamental requisite in
many quantum optics applications \cite{raymer1}. For example, in
the field of linear optical quantum computing (LOQC), the non
fulfillment of these requisites may degrade the quantum gate
fidelity \cite{rohde1}. Various methods to generate single photons
have been proposed and implemented \cite{features}. One of them is
to combine spontaneous parametric down conversion (SPDC) with
conditional measurements: one of the paired photons is used as a
trigger to herald the presence of the other photon
\cite{aichele1}. However, due to the entangled nature of the
photons generated in the SPDC process, the resulting heralded
single photons are not generally described by a pure quantum
state, which severely limits the usefulness of such photons.

The quantum description of photons includes the polarization, the
transverse wave-number distribution and the spectrum. When
considering SPDC sources, indistinguishable paired photons in
polarization can be obtained with a type-I configuration. In the
spatial domain, pure states can be obtained, for instance, by
collecting the downconverted photons with a pair of single-mode
optical fibers.

Regarding the frequency part, pure heralded photons can be
generated if strong spectral filtering is used in the path of the
trigger photon. However, the use of spectral filters represents a
considerable drawback as it results in a loss of the source
brightness, unless the SPDC configuration already generates
narrowband photons, as it is in the case of cavity SPDC
\cite{raymer2,polzik}.

Another way to generate pure heralded photons is to produce
frequency-uncorrelated photons. It has been demonstrated that
frequency-uncorrelated photons generated by SPDC are indeed in a
pure state \cite{uncorrsepara}. Frequency-uncorrelated photons can
be produced in special crystals with suitable pump-light
conditions and specific values of the length and dispersive
properties of the nonlinear crystals \cite{grice1}. Unfortunately,
with this approach the produced photons are not indistinguishable
and one is limited to use specific materials and wavelengths that
can be far from optimal. Various techniques have been proposed to
control the joint spectrum of SPDC pairs. Some methods are based
on the proper preparation of the down-converting crystal
\cite{uren1}; others on the use of angular dispersion to control
the dispersive properties of interacting waves \cite{torres1}; and
others on the use of noncollinear geometries
\cite{uren2,carrasco1}.

\begin{figure}
\centering\includegraphics[scale=0.8,width=1\columnwidth]{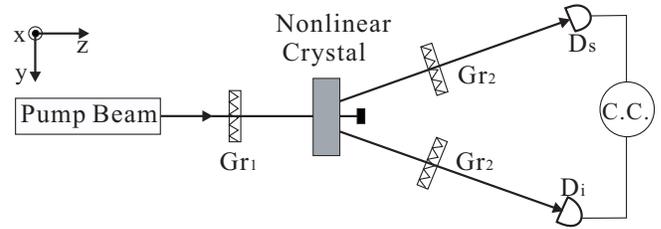}
\caption{General scheme. Gr$_1$ and Gr$_2$: Gratings. D$_s$ and
D$_i$: Single photon counting modules. C.C.: Coincidence counter.}
\label{scheme}
\end{figure}

In this Letter, we propose a new method for tailoring the
frequency properties of SPDC photons that avoids the use of strong
filtering to obtain pure heralded single photons.  In addition,
{\em the technique allows us to tune the frequency bandwidth of the
generated single photons}. This might benefit various
applications, i.e., atom-photon interactions require light with a
narrow bandwidth ($\sim$MHz) while quantum coherence
tomography~\cite{tomogra} and certain quantum information
processing applications~\cite{rohde2} require large bandwidths
($\sim$THz). Importantly, the proposed technique enables to obtain
any kind of frequency correlation between the paired photons
(anticorrelation, uncorrelation or correlation). Compared to other
methods this approach works at any wavelength and for any
nonlinear medium.

The method is based on the fact that the joint spectrum of paired
photons can be independently modified by  a) using noncollinear
geometries that allow mapping the spatial characteristics of the
pump beam into the spectra (spatial-to-spectral
mapping)~\cite{alejandra1} and b) introducing angular dispersion
to modify the group velocities of the interacting fields
(pulse-front-tilt technique)~\cite{hendrych1}.

The scheme is illustrated in Fig.~\ref{scheme}. A noncollinear
degenerate type-I SPDC configuration is used. However, differently
from the standard SPDC, angular dispersion is applied to the pump
beam and the downconverted photons. A diffraction grating or a
prism introduces angular dispersion $\epsilon$ that tilts the
front of the pulse by an angle $\xi$ given by $\tan \xi=-\lambda
\epsilon$, where $\epsilon=m/\left( d \cos \beta_0 \right)$, $m$
is the diffraction order, $d$ the groove spacing, $\beta_0$ the
output diffraction angle and $\lambda$ the wavelength.

The generated signal and idler photons at wavelengths
$\lambda_{s}$ and $\lambda_{i}$ propagate inside the crystal in
the $y$-$z$ plane at an angle $\varphi_s=-\varphi_i=\varphi$ with
respect to the direction of propagation of the pump beam. In
contrast, the angular dispersion is introduced in the orthogonal
$z$-$x$ plane. The diffraction gratings $Gr_2$ in the
down-converted beams compensate for the dispersion introduced by
the grating $Gr_{1}$ in the pump beam, with angular dispersion
$\epsilon^{'}=-\epsilon$.

The quantum state of the SPDC photons writes $|\Psi\rangle=\int
d\omega_s d\omega_i \,\Phi \left(\omega_s,\omega_i\right)
|\omega_s\rangle|\omega_i\rangle$, where $\omega_j$ is the angular
frequency and the subscript $j$ stands for signal ($s$), idler
($i$) and pump ($p$). The two-photon probability amplitude or
biphoton can be written as
\begin{eqnarray}
\Phi \left(\omega_{s}, \omega_{i} \right) & \propto & E_{\omega}
\left(\omega_{s}+ \omega_{i}\right) E_q \left[ \left( k_s-k_i
\right) \sin \varphi\right] \nonumber \\
& \times & {\rm sinc} \left( \frac{\Delta k L}{2} \right) \exp
\left\{ i \frac{\Delta k L}{2} \right\},
\end{eqnarray}
where $E_{\omega}$ is the pump spectrum, $E_{q}$ is the pump transverse momentum
distribution along the $y$ direction and ${\rm sinc}(\Delta k L/2)$ is
the phase matching function. $\Delta k = k_{p}-(k_{s}+k_{i})\cos \varphi$ is the
phase mismatch in the longitudinal direction.

Let us write $\lambda_{j}=\lambda_{j}^{0}+\Lambda_{j}$ where
$\Lambda_{j}$ is the wavelength detuning from the central
wavelength $\lambda_{j}^{0}$. Furthermore, let us define new
variables $\Lambda_{+}=(\Lambda_{s}+\Lambda_{i})/\sqrt{2}$ and
$\Lambda_{-}=(\Lambda_{s}-\Lambda_{i})/\sqrt{2}$ associated with
the diagonal (straight line with a slope of $45^\circ$) and the
anti-diagonal (straight line with a slope of $-45^\circ$) of a two
dimensional density plot of the joint spectrum
$S(\lambda_{s},\lambda_{i}) = \left| \Phi
\left(\lambda_{s},\lambda_{i} \right)\right|^2$ which is the
probability to measure a signal photon with wavelength $\lambda_s$
in coincidence with an idler photon with $\lambda_i$.

The pump spectrum and transverse-momentum amplitude distributions
are assumed to be Gaussian, i.e., $E_{\omega} \left( \omega_p
\right) \propto \exp\left[-\omega_{p}^{2}/(4 B_{p}^{2})\right]$
and $E_{q} \left( \vec{q}_p \right)\propto
\exp\left[-\left|\vec{q}_p \right|^{2}W_{0}^{2}/4\right]$, where
$B_{p}$ is the frequency bandwidth of the pump, $W_{0}$ is the
pump beam waist and $\vec{q}_p=\left( q_x,q_y \right)$ is the
transverse wavevector. Furthermore, we approximate the phase
matching function ${\rm sinc}(\Delta k L/2)$ by an exponential function
of the same width at $1/e$ of the amplitude: ${\rm sinc}(b
x)\simeq \exp[-(\alpha b)^{2}x^{2}]$, with $\alpha=0.455$. If we
project the signal and idler photons into large area spatial
modes, to first order in all frequency variables, the joint
spectrum reduces to
\begin{equation}\label{jointspectrum}
S \left( \Lambda_{s}, \Lambda_{i} \right)={\cal N} \exp \left\{
-\frac{\Lambda_{+}^2}{2 \Delta \Lambda_{+}^2} \right\} \exp \left\{
-\frac{\Lambda_{-}^2}{2 \Delta \Lambda_{-}^2} \right\},
\end{equation}
where $\cal{N}$ is a normalization factor, and $\Delta\Lambda_{+}$
and $\Delta\Lambda_{-}$ are the $rms$ bandwidths of the variable
$\Lambda_{+}$  and $\Lambda_{-}$, respectively, given by
\begin{eqnarray}\label{bplus}
& & \Delta \Lambda_{+}=\frac{\lambda_{s}^{2}}{2\pi c}\frac{1}{\sqrt{2}}
\left[\frac{1}{B_{p}^2}+\left(\alpha L \right)^2 \left(
N'_{p}-N_{s}\cos\varphi \right)^2\right]^{-1/2} \\
& & \label{bminus} \Delta \Lambda_{-}=\frac{\lambda_{s}^{2}}{2 \pi
c}\frac{1}{\sqrt{2}}\left[N_{s} \sin\varphi W_{0}\right]^{-1}.
\end{eqnarray}
$N_{j}= dk_{j}/dw_{j}$ are the inverse group velocities and
$N_{p}^{'}= N_{p}+\tan \rho_{p} \tan\xi/c$ is the effective inverse
group velocity of the pump beam which depends on the Poynting-vector
walk-off angle $\rho_{p}$ and on the pulse-front-tilt angle
$\xi$. $c$ is the speed of light. In all calculations, we consider
typical material parameters corresponding to commonly used
nonlinear crystals such as BBO.

\begin{figure}
\centering\includegraphics[scale=0.9,width=0.8\columnwidth]{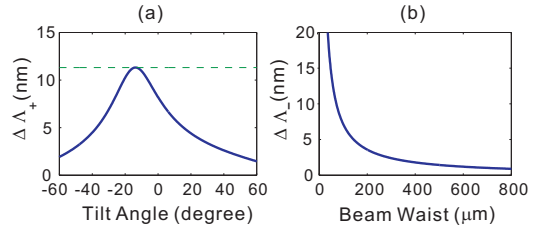}
\caption{Bandwidth of the joint spectrum. (a) $\Delta \Lambda_{+}$
as a function of the pulse-front tilt angle $\xi$. The dashed line
indicates $\Delta \Lambda_{+}^{(max)}$. (b) $\Delta \Lambda_{-}$
as a function of the pump beam waist in the $y$-direction,
$W_{0}$. Pump-beam bandwidth: $\Delta \lambda_p$ = 4 nm.}
\label{bandwidth}
\end{figure}

Eq.~(\ref{jointspectrum}), (\ref{bplus}) and (\ref{bminus}) reveal
the physics behind the proposed technique: Once the pump bandwidth
is fixed, the bandwidth in the $\Lambda_{+}$ direction can be
modified by the pulse-front tilt (see Fig.~\ref{bandwidth}(a)) and
the bandwidth in the $\Lambda_{-}$ direction can be modified by
the size of the pump-beam waist (see Fig.~\ref{bandwidth}(b)).
Eq.~(\ref{bplus}) shows that in the absence of the tilt the
maximum value of $\Delta \Lambda_{+}$ is determined by the
dispersive properties of the material, the length of the crystal
and the noncollinear angle. However, if angular dispersion is
introduced, the phase matching function is modified which allows
us to reach the maximum value $\Delta \Lambda_{+}^{(max)}=2
\sqrt{2}\Delta\lambda_{p}$. This value is achieved by applying a
tilt angle
\begin{equation}
\xi_0=\tan^{-1} \left\{ \frac{c \left( N_{s}\cos\varphi-N_{p}
\right)}{\tan\rho_{p}} \right\}.
\end{equation}

The bandwidth in the $\Lambda_{-}$ direction $\Delta \Lambda_{-}$
can be tailored by changing the pump beam waist at the input face
of the nonlinear crystal in the $y$-direction, $W_0$. This is due
to the so-called spatial-to-spectral mapping that occurs when SPDC
is used in noncollinear geometries~\cite{carrasco1,alejandra1}. In
this configuration, the phase matching conditions inside the
nonlinear crystal enable the mapping of the spatial features of
the pump beam in the $y$-direction into the joint spectrum of the
down-converted photons along the direction $\Lambda_{-}$.

Fig.~\ref{jointspec} shows the joint spectrum for various
combinations of the pulse tilt and the beam waist. Each row
corresponds to a different value of the tilt angle. The first row
depicts the case with no tilt ($\xi=0^\circ$), the second row
corresponds to $\xi=\xi_0=-13.8^\circ$, which yields the maximum
bandwidth in the $\Delta \Lambda_{+}$ direction that can be
obtained for a given pump bandwidth. The third row corresponds to
$\xi=30^\circ$. When the bandwidths in the $\Lambda_{+}$ and
$\Lambda_{-}$ directions are equal, indistinguishable, and
frequency-uncorrelated photons are generated. This can be achieved
by choosing an appropriate combination of the tilt and the beam
waist as depicted in the central column of the figure. It can be
easily seen how by modifying the tilt and the beam waist, the
frequency bandwidth of single photons can be modified.

Fig.~\ref{jointspec} also reveals that the setup discussed for the
generation of pure heralded photons allows the production of
paired photons with different types of frequency correlations. As
a matter of fact, frequency uncorrelation is just a particular
case. The first and third column of Fig.~\ref{jointspec}
correspond to different values of the pump beam waist in the
$y$-direction, $W_0=30$~$\mu$m and $W_0=250$~$\mu$m, respectively.
The first column depicts highly frequency-anticorrelated photons,
while the third column illustrates the case of highly
frequency-correlated photons.

\begin{figure}
\centering\includegraphics[scale=0.9,width=0.9\columnwidth]{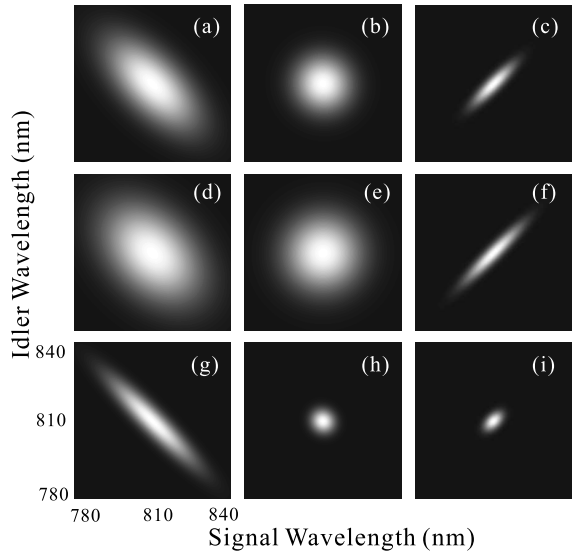}
\caption{Joint spectrum of the two-photon state for different
values of the pulse-front tilt ($\xi$) and the beam waist
($W_{0}$). (a) $\xi=0^\circ$ and $W_0=30$~$\mu$m; (b)
$\xi=0^\circ$, $W_0=60$~$\mu$m; (c) $\xi=0^\circ$,
$W_0=250$~$\mu$m; (d) $\xi=\xi_0=-13.8^\circ$, $W_0=30$~$\mu$m;
(e) $\xi=\xi_0=-13.8^\circ$, $W_0=45$~$\mu$m, (f)
$\xi=\xi_0=-13.8^\circ$, $W_0=250$~$\mu$m; (g) $\xi=30^\circ$,
$W_0=30$~$\mu$m; (h) $\xi=30^\circ$, $W_0=140$~$\mu$m, and (i)
$\xi=30^\circ$, $W_0=250$~$\mu$m. The circular shape of the
distributions shown in the central coulumn indicates frequency
uncorrelation between the photons with different bandwidth.}
\label{jointspec}
\end{figure}

In conclusion, a new technique for the generation of heralded
indistinguishable and pure single photons with a tunable frequency
bandwidth has been presented. The full control of the joint
spectrum allows us to generate frequency-correlated and
frequency-anticorrelated photon pairs as well. The proposed method
combines SPDC in noncollinear geometries with the use of
pulse-front tilt. The control parameters used to tune the
frequency characteristics are readily accessible experimentally:
pump beam width and angular dispersion. The method described here
works in any frequency band of interest and does not require any
specific engineering of the dispersive and nonlinear properties of
the nonlinear medium.

Acknowledgements: This work has been supported by the European
Commission (QAP, IST directorate, Contract No. 015848), and by the
Government of Spain (Consolider Ingenio 2010 QIOT CSD2006-00019
and FIS2007-60179). MH acknowledges support from a Beatriu de
Pinos fellowship.

\newpage

\end{document}